\documentclass[onecolumn]{aastex6}

\collaborationName{Friends of AASTeX}
\begin{document}
\title{High-resolution Observation of Downflows at One End of a Pre-eruption Filament}
\author{Qin Li\altaffilmark{1}, Na Deng\altaffilmark{1,2}, Ju Jing\altaffilmark{1,2}, and Haimin Wang\altaffilmark{1,2}}
\altaffiltext{1}{Space Weather Research Laboratory, New Jersey Institute of Technology, 
University Heights, Newark, NJ07102-1982, USA;ql47@njit.edu 
}
\altaffiltext{2}{Big Bear Solar Observatory, New Jersey Institute of Technology, Big Bear City, CA 92314-9672, USA}

\begin{abstract}

Studying the dynamics of filaments at pre-eruption phase can shed light on the precursor of eruptive events. Such studies in high-resolution (in the order of 0.1") are highly desirable yet very rare so far. In this work, we present a detailed observation of a pre-eruption evolution of a filament obtained by the 1.6 m New Solar Telescope (NST) at Big Bear Solar Observatory (BBSO). One end of the filament is anchored at the sunspot in NOAA active region (AR) 11515, which is well observed by NST H$\alpha$ off-bands four hours before till one hour after the filament eruption. A M1.6 flare is associated with the eruption. We observed persistent downflowing materials along the H$\alpha$ multi-threaded component of the loop towards the AR end during the pre-eruption phase. We traced the trajectories of plasma blobs along the H$\alpha$ threads and obtained the plane-of-sky velocity of 45 km s$^{-1}$ on average. We further estimated the real velocities of the downflows and the altitude of the filament by matching the observed H$\alpha$ threads with magnetic field lines extrapolated from a nonlinear force-free field (NLFFF) model. Observation of chromospheric brightenings (BZs) at the footpoints of the falling plasma blobs is also presented in the paper. The lower limit of the kinetic energy per second of the downflows through the BZs is found to be $\sim$ 10$^{21}$ erg. Larger FOV observations from BBSO full disk H$\alpha$ images show that the AR end of the filament started ascending four hours before the flare. We attribute the observed downflows at the AR end of the filament to the draining effect of the filament rising prior to its eruption. During the slow-rise phase, the downflows continuously drained away $\sim$ 10$^{15}$g mass from the filament over a few hours, which is believed to be essential for the instability at last, and could be an important precursor of eruptive events. 

\end{abstract}
\section{Introduction} \label{sec:intro}	
Solar filaments (on disk)/prominences (on limb) have been studied for decades	 and their structures have become one of the well known features on solar atmosphere. They are formed of numerous cool and dense plasma threads upheld by highly sheared and twisted magnetic fields in chromosphere and corona, and can last for days or even weeks in corona until disappear in-situ or violent eruption \citep{1973SoPh...31....3M}. Moreover, filament eruptions could be part of other dynamic and energetic events such as flares and coronal mass ejections (CMEs) \citep{2012SoPh..281..237V}. Filament evolution can give insight to the local magnetic field structure which otherwise cannot be measured directly. For instance, filaments' spatial evolution discloses the surrounding magnetic field topologies as a flare evolves \citep{1974SoPh...34..323H}.

%At low temperature, partially ionized plasma is confined by magnetic fields, and the field topology guides the motion of trapped matters. On the other hand, counter-streaming mass flow along interleaved threads of filaments was first reported by \citet{1998Natur.396..440Z}, which suggests the ubiquitous existence of bidirectional mass streaming and vertical fields in quiet prominence. Also visible vertical threads in quiet prominences are able to be observed in EUV, such as AIA 193 $\AA$ off-limb \citep{2015SoPh..290.1703M}, in the form of hedgerow topologies. These constant gravitational drainages \citep{2010ApJ...716.1288B} are parts of the coronal cycle, being balanced out by ubiquitous upward flows, and highly linked to filament formation and maintenance mechanisms. 

Filament is composed of spine, barbs and two ends. They are lying on sheared arcade and above magnetic polarity inversion line  (PIL) \citep{2015ApJ...806....9K}. According to some models, the structure of dipped field lines are responsible for dense matter accumulation \citep{1989A&A...221..326D}. Such dip structures form in the magnetic configuration allowing dense matters accumulate and tend to bend down at the top of arcades \citep{1998A&A...329.1125A}. Filament naturally involve radial outward magnetic force, so the weight of these dense matters balance out the upward Lorentz force, and result in the filament equilibrium. Magnetic structure expulsion therefore was believed to be a consequence of not capable of confining the magnetic fields \citep{2002ApJ...571..987F}. During the eruption downflows are expected to be present to drain away the most part of filament material, and accompanied with the upward transport of magnetic flux due to the magnetic fields ascent  \citep{2005ApJS..159..288P}. Draining rate of 10$^{15}$ g in a quiescent filament per day was estimated  by \citet{2012ApJ...745L..21L}. The magnetic energy that is required for the eruption could be as large as the weight of removal materials that holds the entire magnetic structure in force equilibrium. 

Filament dynamics have been studied in H$\alpha$ and EUV observations through the analysis of  the paths of absorption features and Doppler shift \citep{2014IAUS..300...69P}. Tracing features of time sequence images and analyzing spectral line profiles obtain two complementary components: plane-of-sky velocity and line-of-sight (LOS) motion of mass flows. Counter-Streaming mass flow along interleaved threads of filaments was first observed through H$\alpha$ Doppler map with a speed of 5$\sim$20 km s$^{-1}$\citep{1998Natur.396..440Z}. Individual barbs stretched from a quiescent filament spine in Ca II H and H$\alpha$ lines show the horizontal speed of 10 km s$^{-1}$ and the downward speed of 35 km s$^{-1}$ \citep{2015ASSL..415...79K}. The mass flow of 25 - 30 km s$^{-1}$ plane-of sky velocity from the spine that led to bard formation was also reported by \citet{2013SoPh..288..191J}. In earlier review by \citep{2010SSRv..151..333M}, downflow blobs are moving along the vertical threads in $\sim$10 km s$^{-1}$. These flows have velocities on average lower than sound speed in 20-40 km s$^{-1}$ for prominence plasma \citep{2014LRSP...11....1P}. But in EUV observation, the flow motion in a filament is faster that in H$\alpha$ observation, e.g., the counter-steaming flow motion in an active filament can reach to 100 km s$^{-1}$ in EUV 193 $\AA$ band \citep{2013ApJ...775L..32A}. As suggested by \citet{2015ASSL..415...79K}, EUV observations of the moving plasma features have shown faster than what is considered normal in H$\alpha$. This is because that the hot plasma as observed in EUV is not visible in H$\alpha$. 

Threads of filament mass flow are to some extent analogue to coronal rain, the scene of which could be described as plasma moving along the magnetic field lines \citep{2012SoPh..280..457A}. The mechanism accountable for filament formation also resembles that of coronal rain due to condensation yet in different rate. Filament mass is maintained by condensation at a high rate of 10$^{10}$ g s$^{-1}$ \citep{2012ApJ...745L..21L}. In coronal rain the temperature transition of coronal loops along their height and the falling sunward blobs could be observed through AIA 171 $\AA$ and 304 $\AA$ off-limb observation\citep{2015A&A...577A.136V}. Compared to coronal rain blobs which are relatively loose and faint, those threads of filaments downflows are much more continuous and denser. 

Filament drainage is a common phenomenon either in quiescent phase or in eruption phase. They are parts of the coronal cycle  \citep{2010ApJ...716.1288B}, being balanced by the ubiquitous upflows and condensation. Their speed can be as low as 5 km s$^{-1}$ \citep{2014IAUS..300...69P}. strong draining can be observed in eruption phase. \citet{2014A&A...566A.148H} observed the filament drainage in pre-eruption phase. Such drainage in pre-eruption phase has also been reported by \citet{2014ApJ...790..100B}. Large-amplitude filament oscillation was observed in pre-eruption phase, and followed by the substantial material drainage prior to eruption. The authors suggested that drainage could play a role in the slow-to-fast transition of filament eruption.

\begin{figure}
\figurenum{1}
 \gridline{\fig{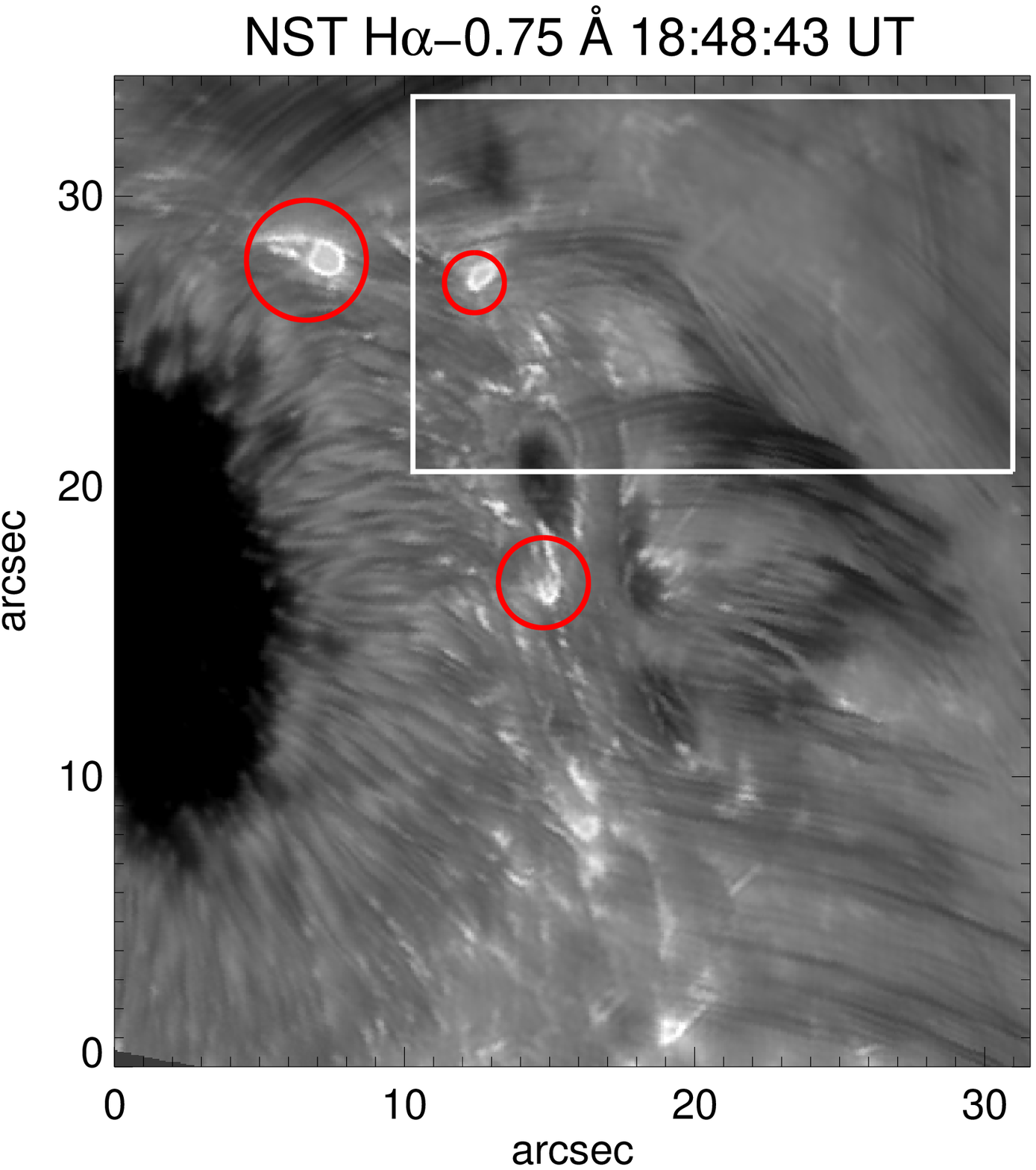}{0.4\textwidth}{(a)} 
 \fig{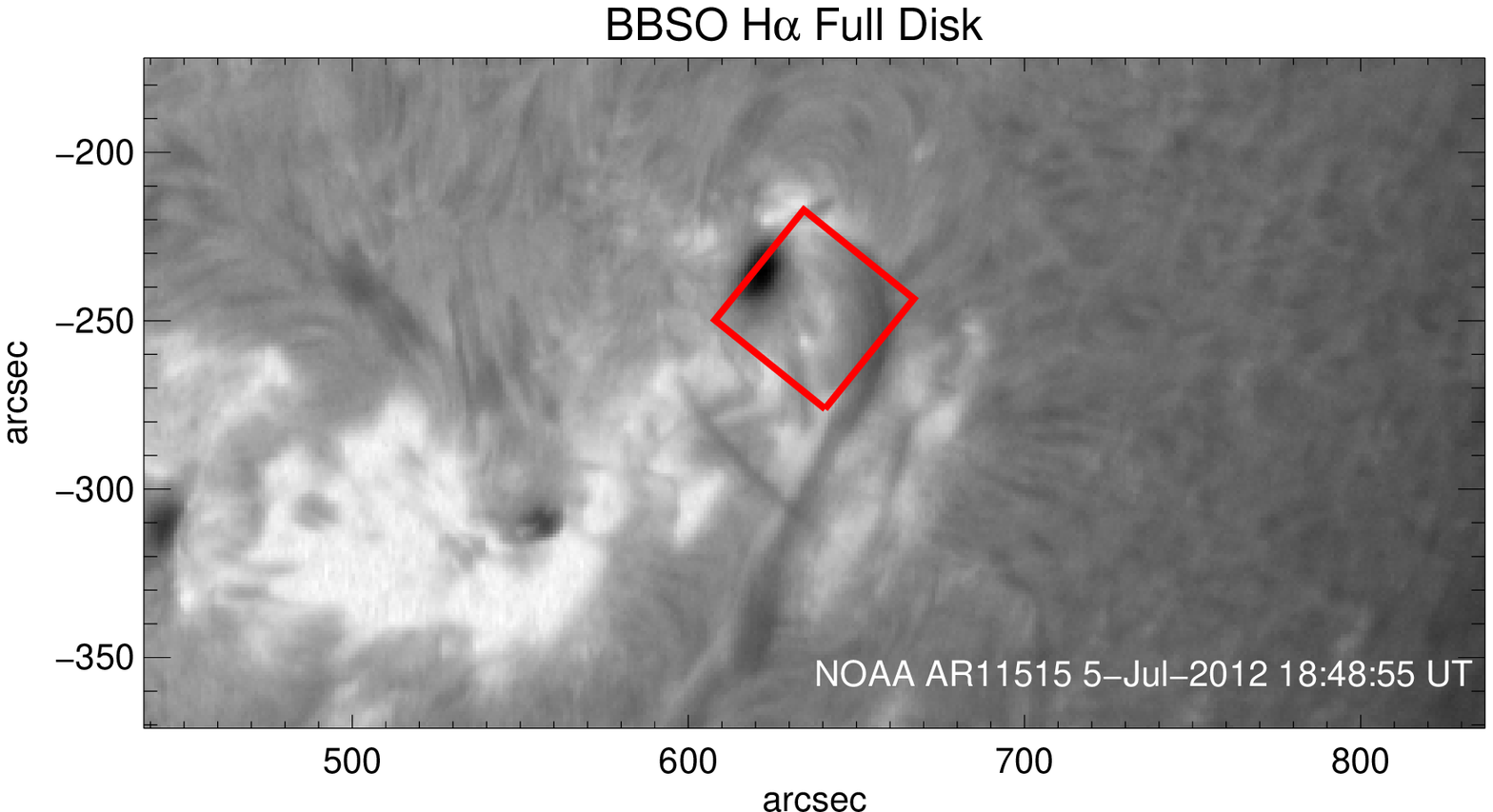}{0.6\textwidth}{(b)}  }
\gridline{\fig{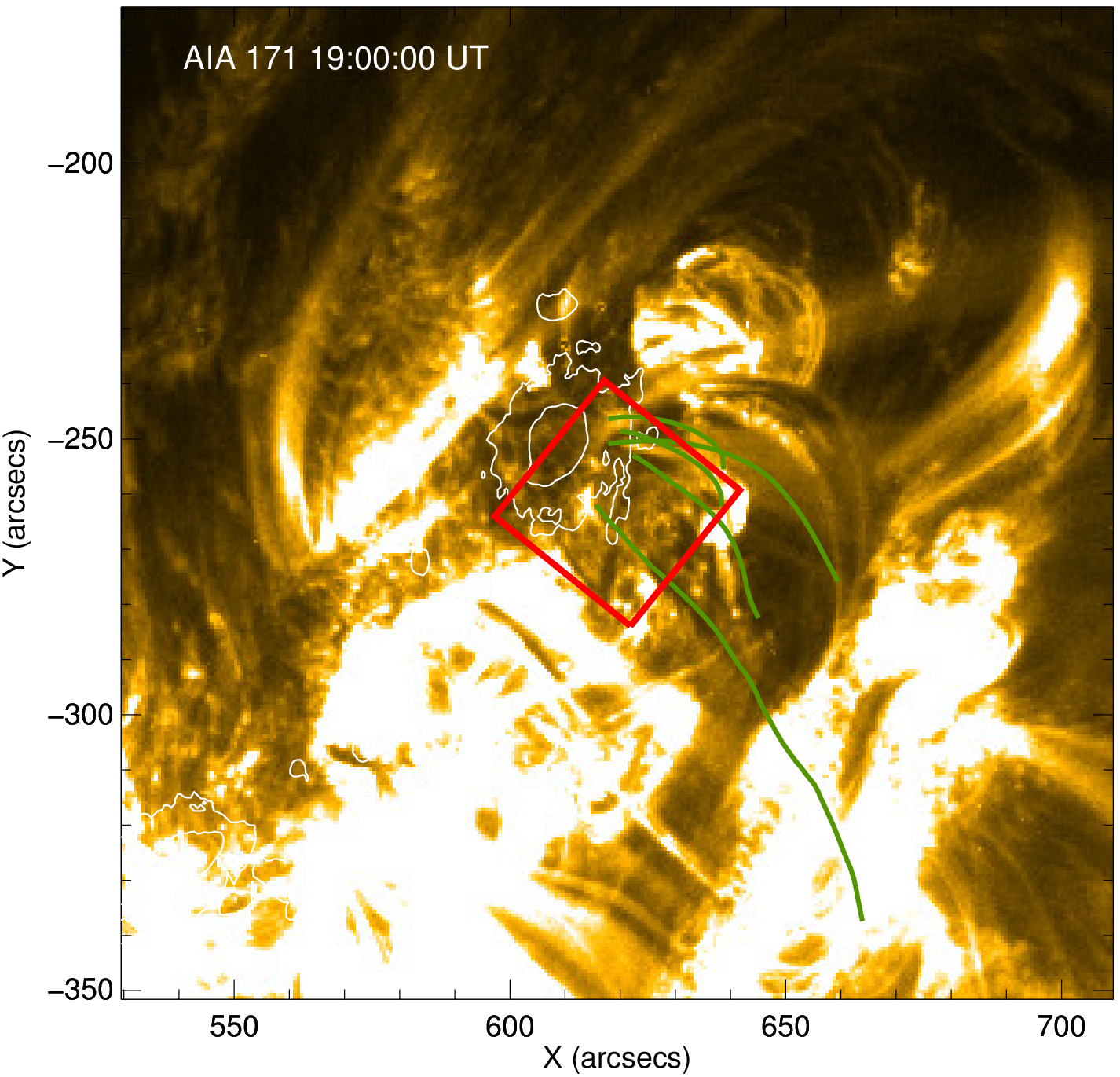}{0.5\textwidth}{(c)} 
	 \fig{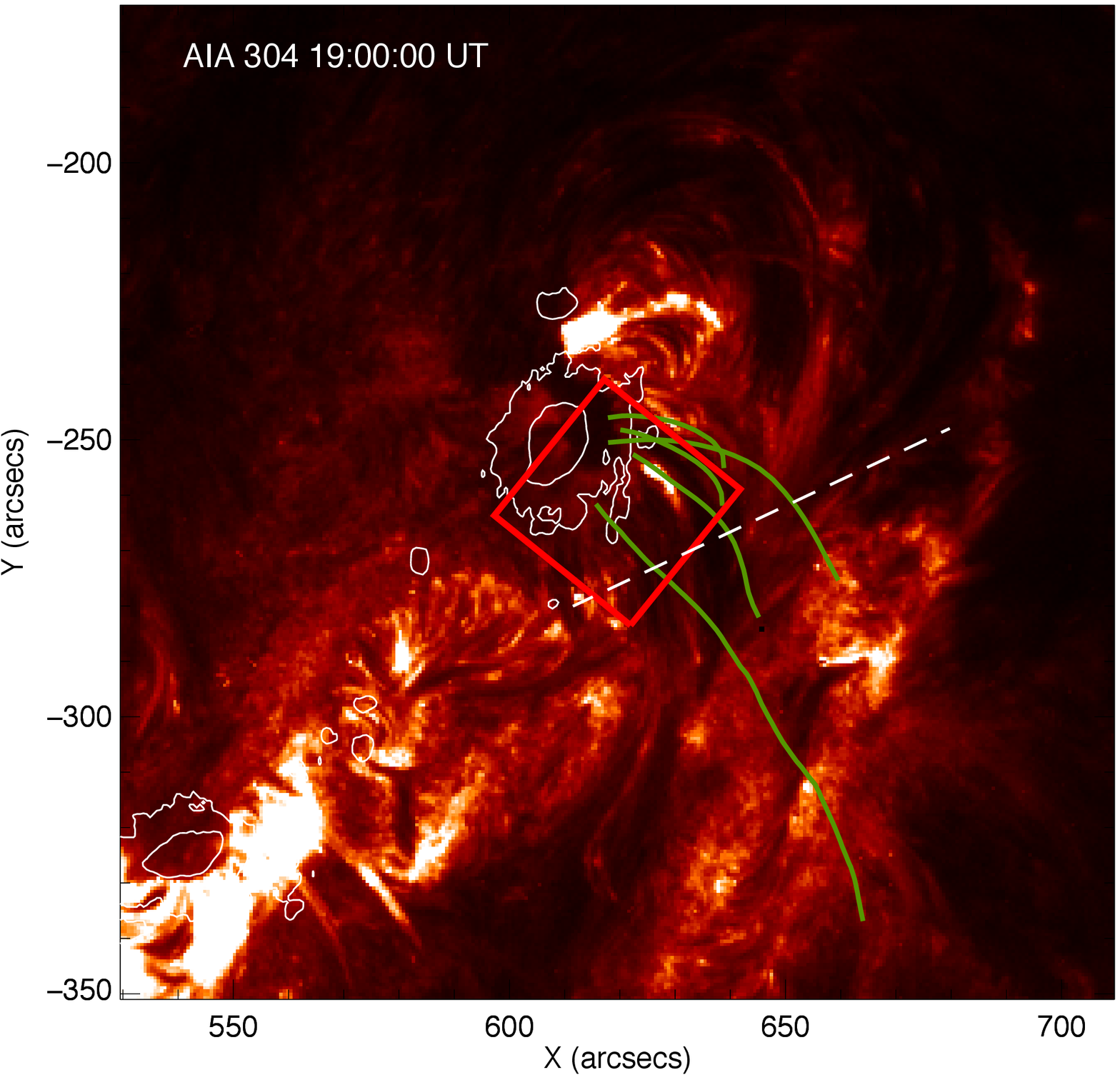}{0.5\textwidth}{(d)}
          }
\caption{Location of the filament as it is seen one end of filament at the AR 11515 on July 5, 2012. (a) NST high-resolution image, the white box indicates the FOV within which downflow thread dynamics is studied and shown in Figure 2, the red circles outline some chromospheric brightenings, (b)H$\alpha$ full disk image, the red box shows the FOV of NST observation shown in (a),  (c) AIA 171 $\AA$ image, (d) AIA 304 $\AA$ image. White contours outline the sunspot, the red box shows the FOV of NST observation, the dashed line marks the trajectory for the slit-image shown in Figure 7 left panel, and the green lines show modeled field lines from NLFFF extrapolation. An animation of NST high-resolution observations is available online.\label{fig:pyramid}}
\end{figure}
 
%Lack of high spatial resolution will lead to the difficulties in observing them. Doppler measurement is used to derive dense material's LOS motions, i.e. Coronal loops' dynamics \citep{2008AnGeo..26.2955D}, filament countering streaming and filament ascent \citep{2015ApJ...806....9K}. But without high-resolution spectroscopic imaging, it is difficult to determine the sources of the redshift and blueshift spectra, i.e, spatial difference between upward and downward  flows. When the use of line profiles combines with the high resolution imaging, both LOS and apparent dynamics can be attained for understanding.While many studies have been come out for quiet-sun filament, such as the downflows, vortices and oscillations dynamics\citep{2008ApJ...676L..89B}, active filaments are not well studied especially with high spatial resolution spectroscopy. The difficulty for studying fine-structure filament dynamics is that lack of high resolution observational instruments in an order of 0.1" spatial resolution.

 The fine-structure (\textless 500 km) filament drainage can help to understand the eruption mechanism, yet have been rarely observed due to lack of high spatial resolution observation. In this paper, the high-resolution H$\alpha$ data are from the 1.6 m New Solar Telescope (NST) at the Big Bear Solar Observatory (BBSO) that achieves 0.1" spatial resolution. We observe one end of a small filament in an AR. During the five hour observation period, we see the filament ascending and strong consequential downflows towards this end. Unfortunately LOS velocity from doppler measurement cannot be deduced due to lack of H$\alpha$ red wing observation, but we can still find out the characteristics of the strong downflows before eruption by considering the local magnetic field topology and identify the potential role of the drainage as a pre-eruption precursor. 

\begin{figure}
\figurenum{2}
\gridline{\fig{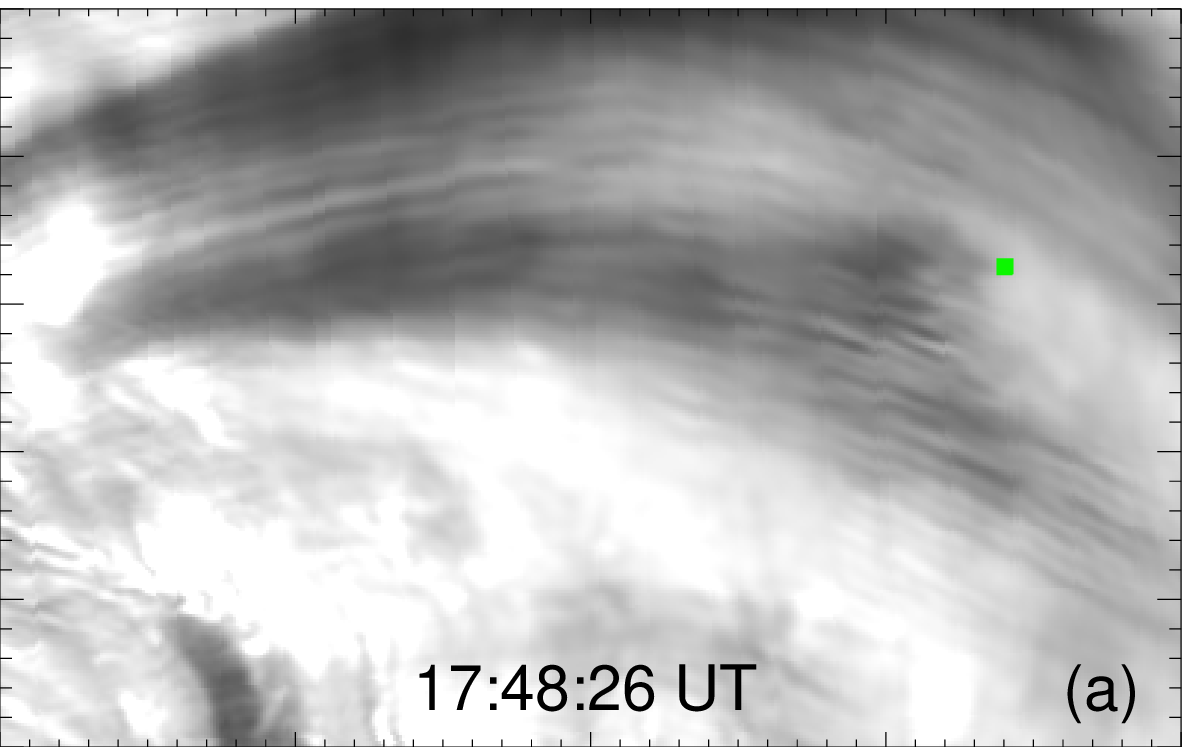}{0.25\textwidth}{}
          \fig{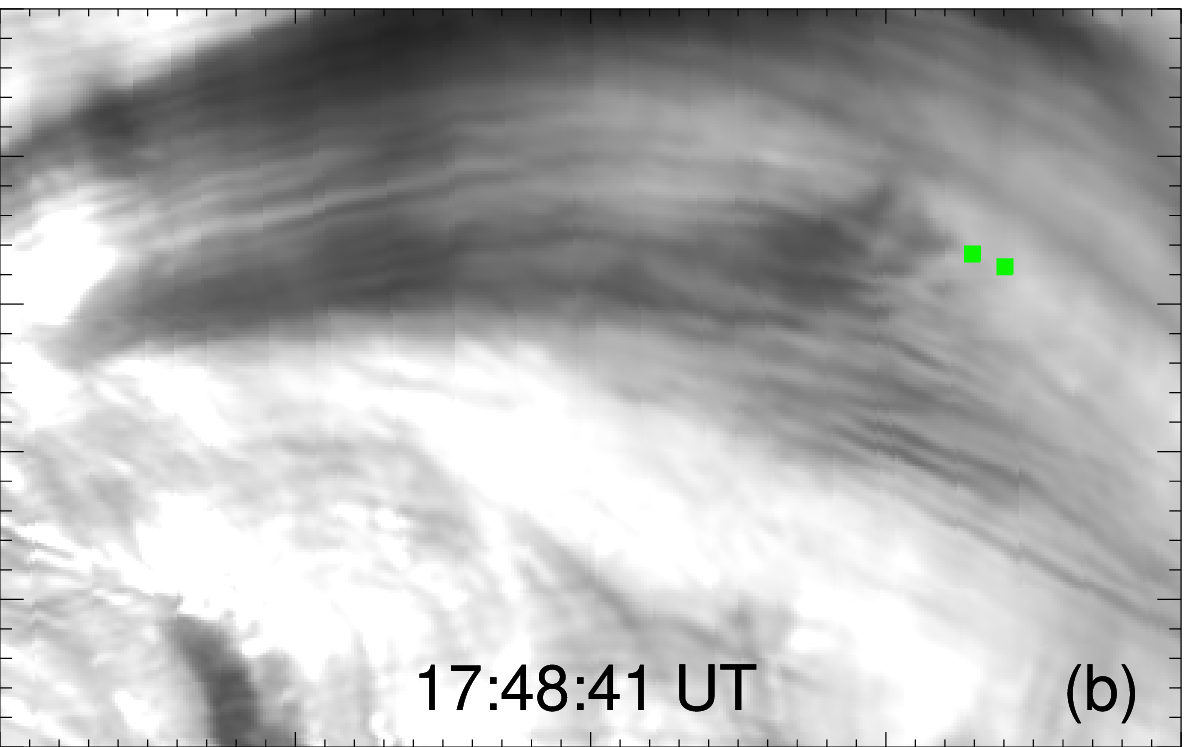}{0.25\textwidth}{}    
          \fig{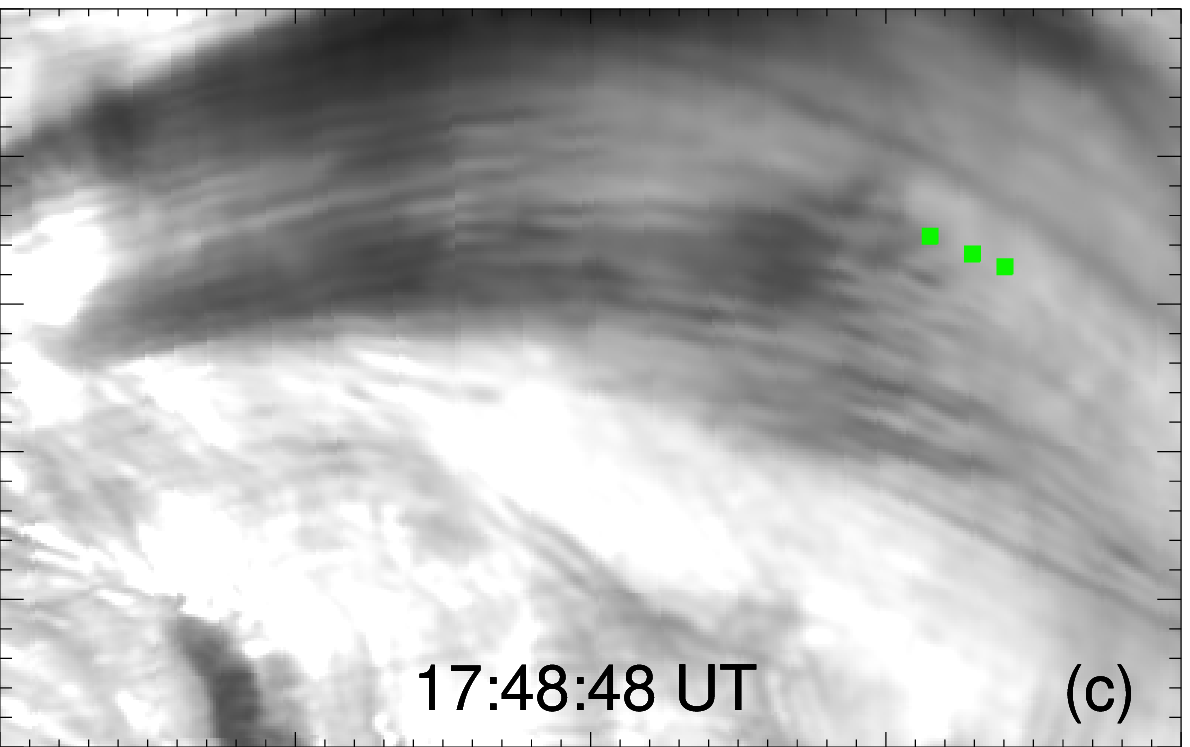}{0.25\textwidth}{}  
          \fig{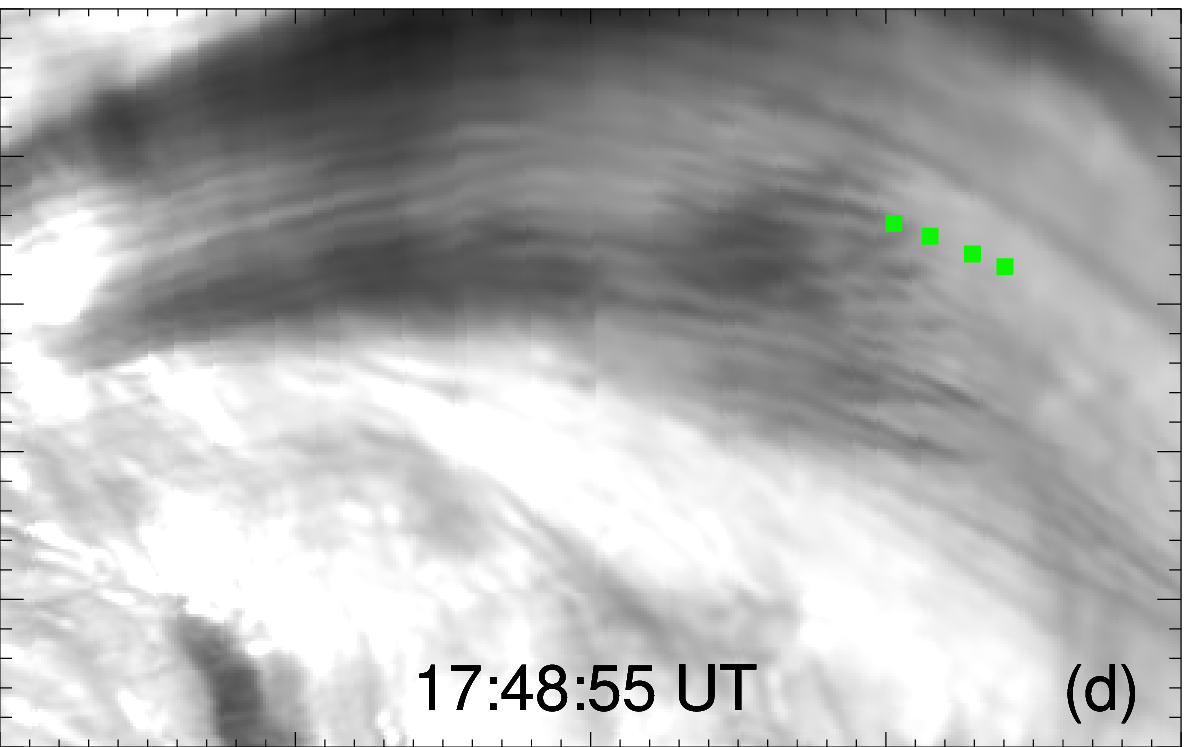}{0.25\textwidth}{}  
          }
\gridline{\fig{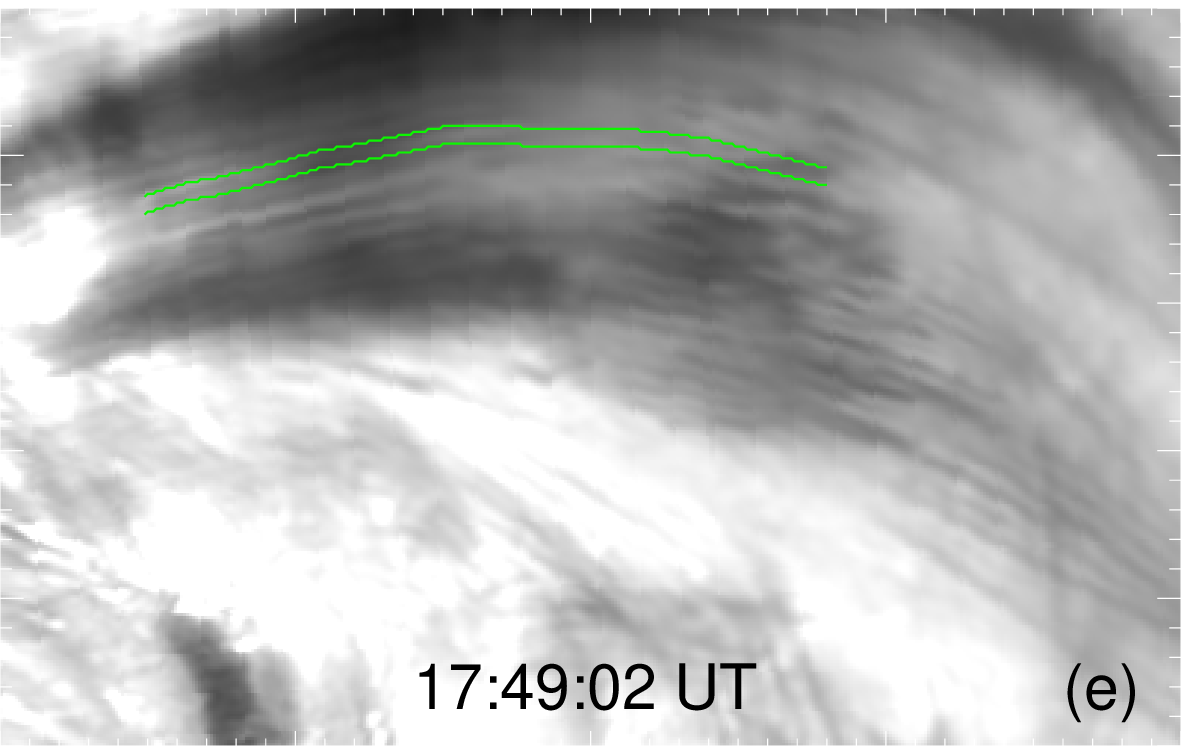}{0.25\textwidth}{}
	\fig{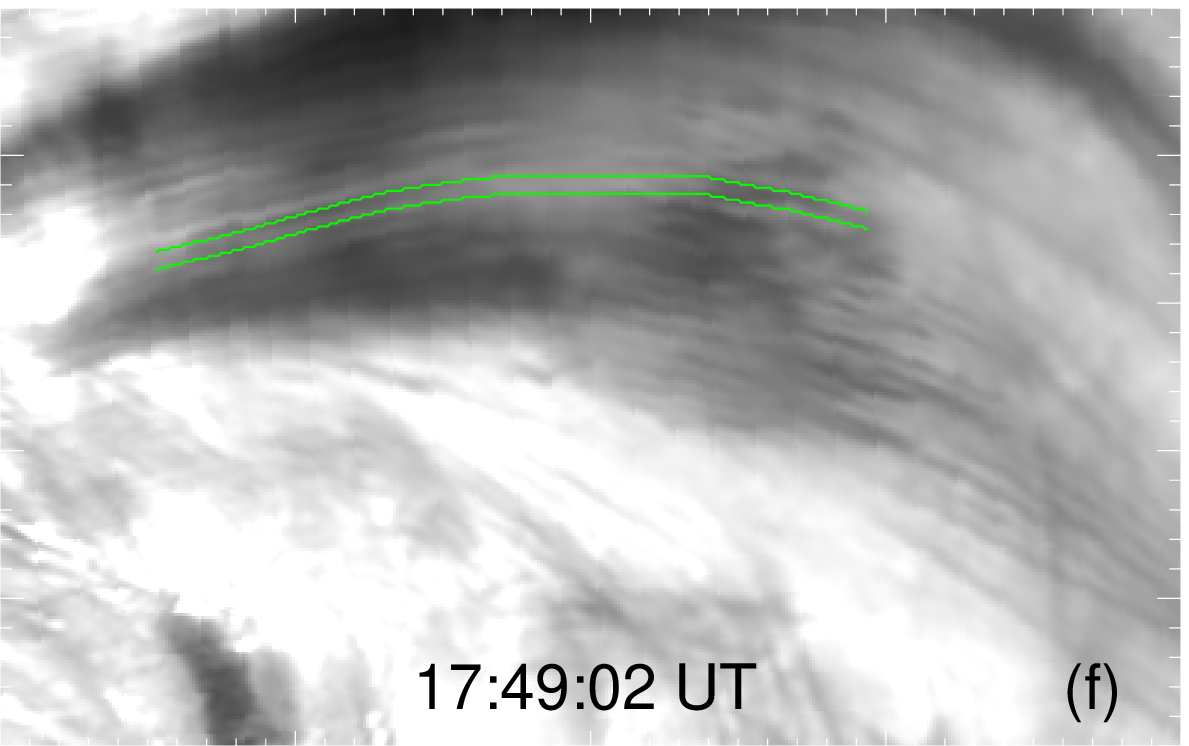}{0.25\textwidth}{}
	\fig{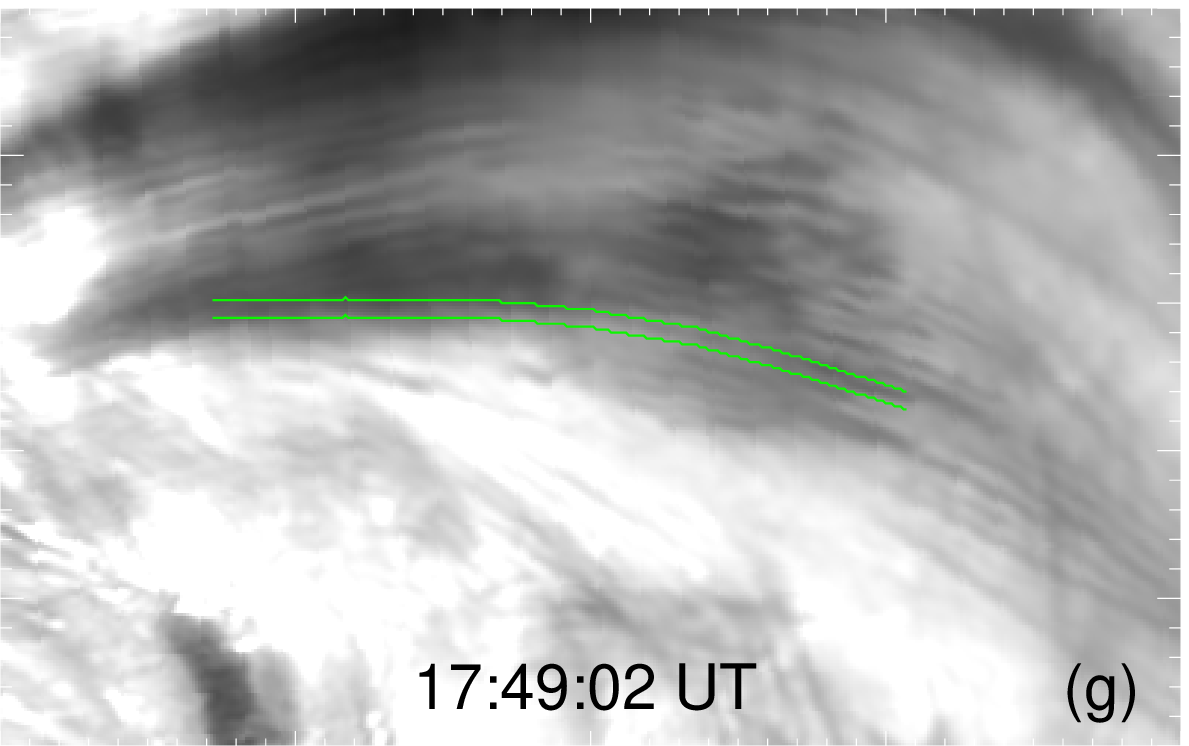}{0.25\textwidth}{}
	\fig{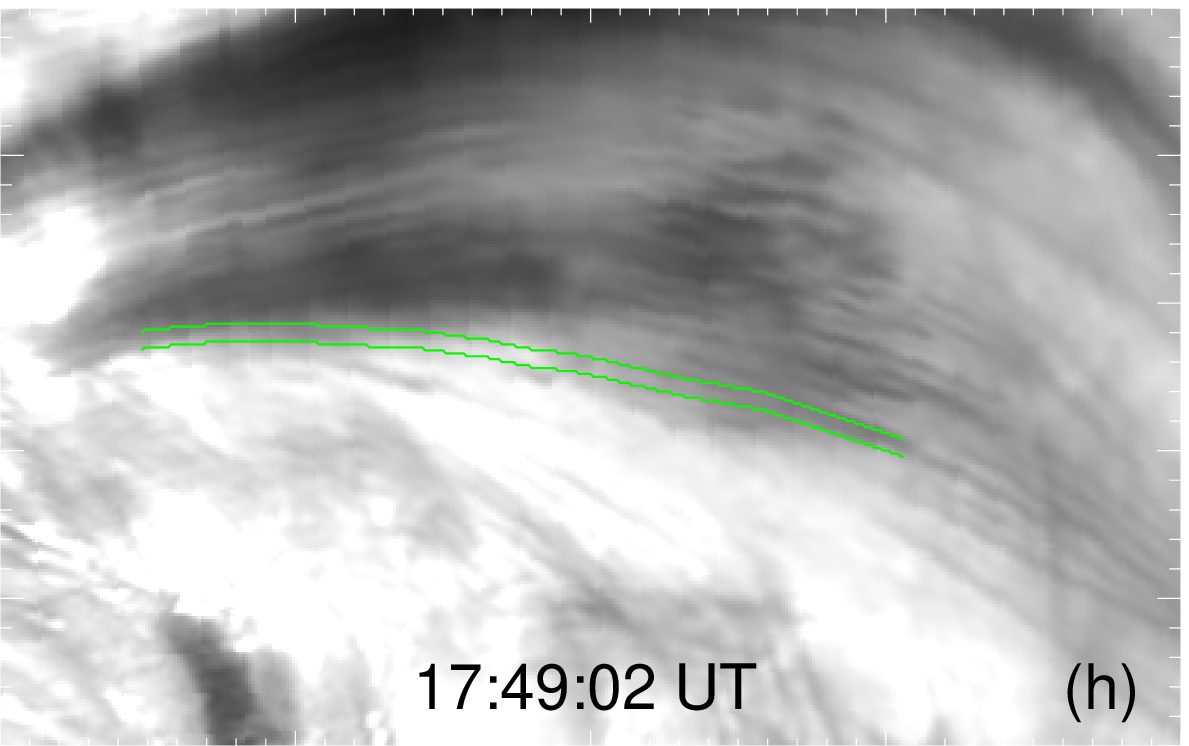}{0.25\textwidth}{}
	}
\gridline{\fig{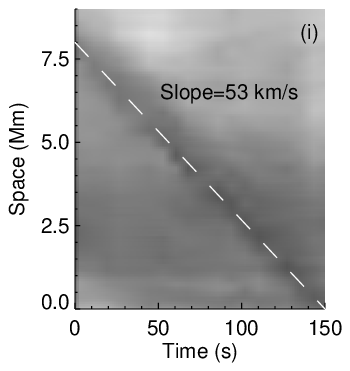}{0.25\textwidth}{}
	 	\fig{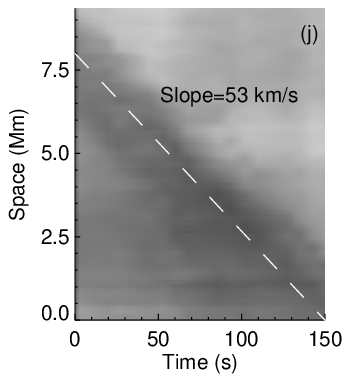}{0.25\textwidth}{}
		\fig{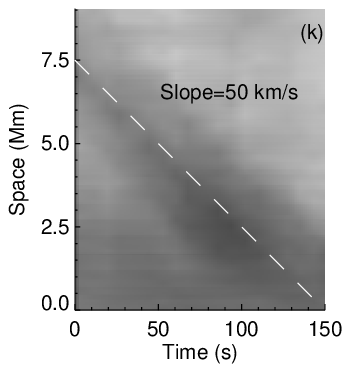}{0.25\textwidth}{}
		\fig{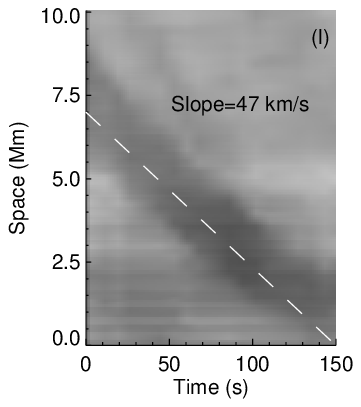}{0.25\textwidth}{}
	 }
\caption{(a - d) show blobs tracing method by marking nodes along downflow thread trajectory. (e - h) show four different trajectories as outlined by green curves over the same time period in progressively lower altitude. The green curves also denote the slits used to derive the time-distance diagrams displayed in (i - l). (i - l) show the respective space-time diagrams of the four curved trajectories from green slits in (e - h). Dashed lines are fitting results to obtain average velocities of the blobs. \label{fig:pyramid}}
\end{figure}

\section{Observation and data description} \label{sec:style}

The main data source is H$\alpha$ images from the Visible Imaging Spectrometer (VIS) at NST. In earlier years of its operation,  NST was operated with the 76-element Adaptive Optics System\citep{2010ApJ...714L..31G}. In the visible wavelengths, NST achieved a spatial resolution around 0.05" after using 100 frames for speckle reconstruction. The cadence of data is 15 sec.  On 2012 July 5,  the H$\alpha$ observations were made through a tunable filter at two wavelengths:    
 H$\alpha$ line center (=6562.808 $\AA$) and 0.75 $\AA$ off band in the blue wing.  NST has a limited FOV around $50" \times 50"$, therefore it only covers a fraction of the targeted AR 11515.  In addition, observations are also made in TiO band (a proxy for the continuum photosphere at 7057 $\AA$).  The full-disk H$\alpha$ line center images were also obtained at BBSO with a separate telescope with a cadence of  one minute, and a pixel size in the order of 1''. A small AR filament is visible, one end of the filament is located in NST FOV at the edge of  the largest sunspot of the region (see Figure 1). It is worth noting that a M1.6 flare occurred at 21:39 UT,  covered by NST observations.

 We also analyze the EUV images obtained from the AIA on board Solar Dynamic Observatory(SDO), in particular, the EUV images at the wavelength of 171 $\AA$ and 304 $\AA$, which correspond to the corona and upper chromosphere of the Sun, respectively.  The 171 $\AA$ band is mainly from Fe IX line, with the characteristic temperature of 10$^{5.8}$ K, while  the 304 $\AA$ is HeII line, with the characteristic temperature of 10$^{4.7}$ K \citep{2012SoPh..275...17L}. 

To understand the magnetic structure, we analyzed the vector magnetograph data from HMI on board SDO by conducting NLFFF extrapolation. Non-linear force free field (NLFFF) extrapolation model assumes that the corona is static and free of Lorenz Force: J$\times$B=0. And the currents must be aligned with the magnetic field: $\bigtriangledown$$\times$B=$\alpha$B, where $\alpha$ is the non linear coefficient. Under NLFFF approximation, by assuming the Lorentz forces negligible, $\alpha$ varies in  space, and field evolves slowly through a series of quasi-equilibrium states \citep{2008SoPh..247..269M}

We used $\verb#hmi.sharp_cea_720s#$ series data which are disambiguated vector magnetograms with a pixel size of 0.5\arcsec. It converts data as ($B_r$,$B_\theta$,$B_\phi$) in a heliocentric spherical coordinate corresponding to ($B_z$,-$B_y$,$B_x$) in the heliographic coordinates \citep{2013arXiv1309.2392S} through cylindrical equal area (CEA) projection. We congrid the magnetograms to 2.25\arcsec\ pixel and preprocess the data toward a suitable boundary condition to achieve nearly force-free conditions \citep{2006SoPh..233..215W}. Optimization code with weighting function was also conducted \citep{2004SoPh..219...87W}. The extrapolation was performed within a computational domain of 256$\times$100$\times$200 uniform grid points, corresponding to $\sim$450$\times$175$\times$350 Mm$^{3}$.

The most important part of data processing is to trace flow speed of plasma motions based on H$\alpha$ observation.  We  assume that the plasma motion follows magnetic field lines.  This is generally true when plasma pressure is much less than magnetic pressure, The detailed procedure to track the flows including the correction of projection effect will be discussed in the next section. 

\begin{figure}[ht!]
\figurenum{3}
\plotone{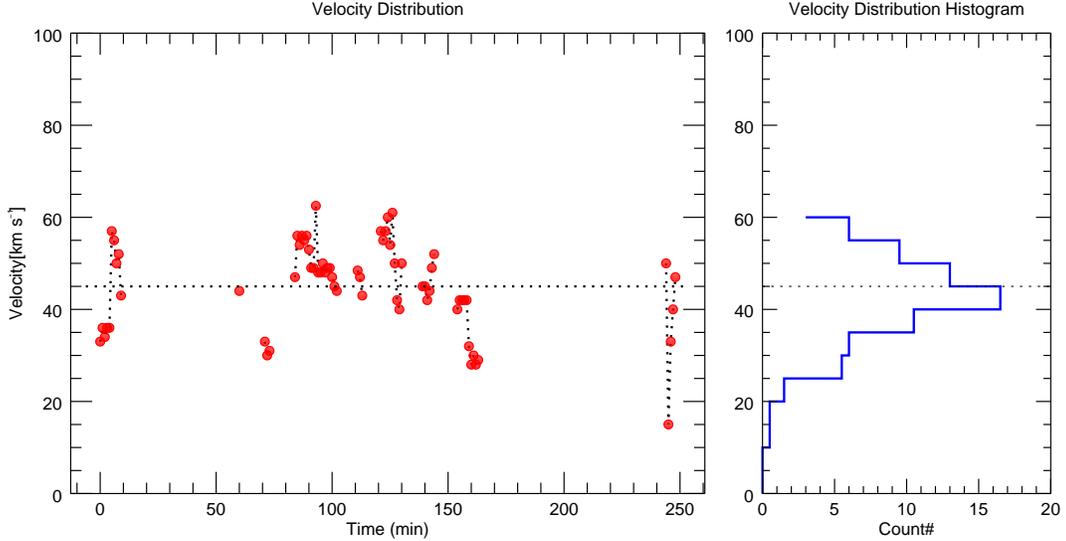}
\caption{The statistical distribution of detectable blobs' plane-of-sky velocities and the corresponding plane-of-sky velocity histogram. Dashed horizontal line indicates average velocity of downflow threads.\label{fig:general}}
\end{figure}

\section{Results} \label{sec:floats}
\subsection{Downflowing Threads at the Filament End} \label{subsec:tables}
Figure 1 shows the filament in NOAA AR 11515 on July 05, 2012, as seen in H$\alpha$ (Figure 1(a)(b)), AIA 171 \AA\ (Figure 1(c)) and 304 \AA\ (Figure 1(d)). In particular, the FOV of the NST/VIS high-resolution H$\alpha$ observation covers one end of the filament (Figure 1(a)) from 16:41 to 22:22 UT. The filament eruption was associated with a flare at 21:39:00UT. In Figure 1(c)(d), the filament is seen in AIA 304 $\AA$ as a dense loop consisting of dark materials, which shows the low temperature property of the filament. Rapid temperature transition from 1 MK to 0.05 MK was observed by \citet{2015A&A...577A.136V}. For this event, however, no clear temperature transition can be determined, which may imply a different driver mechanism. The sunspot contours are obtained from HMI white light images. The downflow trajectories are marked in green lines. NST data provides details of filament threads (white box) and green lines' footpoints (red circles) (Figure 1(a)). The FOV of NST data is marked in red in Figure 1(b). We observe persistent downflows along the filament threads four hours before the eruption. 
In Figure 2, We used the moving plasma blobs as a tracer of the downflow threads, and tracked the trajectories of the head (or tail) of these blobs as shown in Figure 2 (a -- h). In this example, about 15 points in average are pinned to determine each trajectory over a period of $\sim$150 seconds. The trajectories of four sample downflow threads are shown in Figure 2 (e -- h).

\begin{figure}
\figurenum{4}
\gridline{ \fig{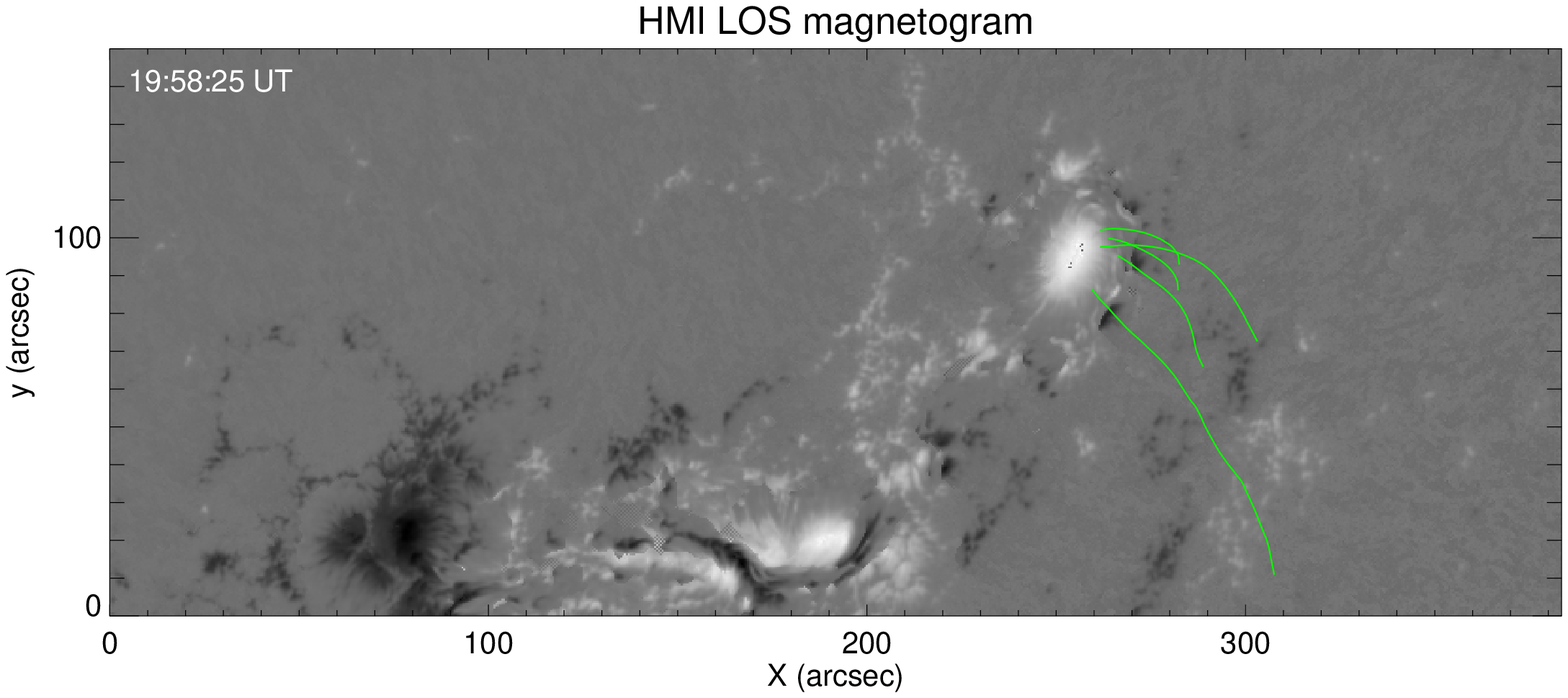}{0.6\textwidth}{(a)}
          \fig{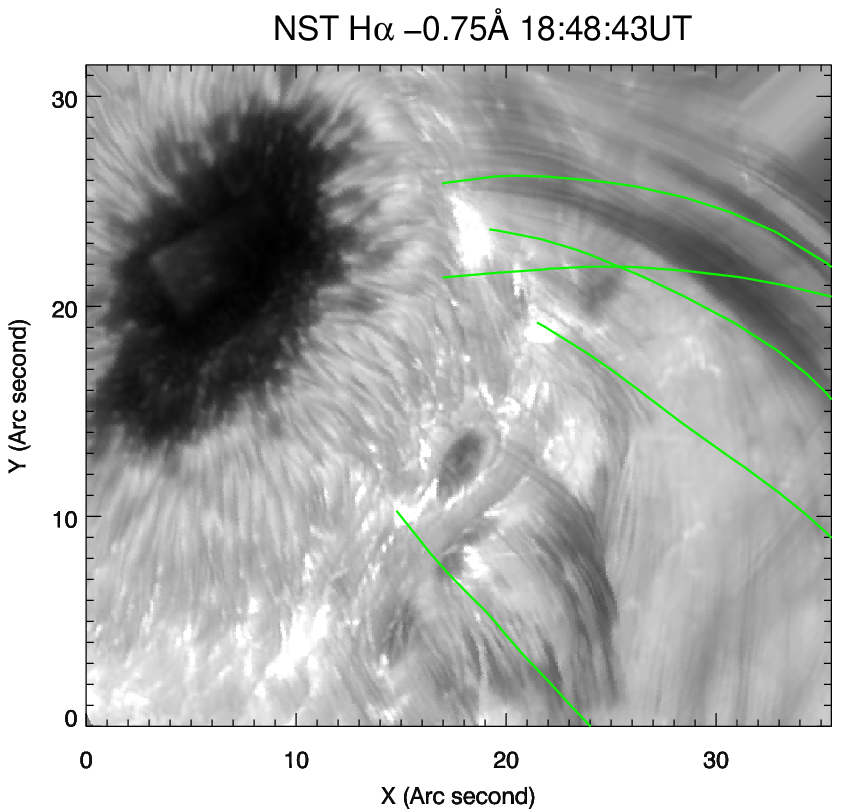}{0.4\textwidth}{(b)}
	}
\caption{Magnetic field lines (green lines) are extrapolated from the NLFFF model. (a) Magnetic field lines extrapolated from five footpoints of downflows, (b) Extrapolated field lines plotted on an H$\alpha$ image. \label{fig:pyramid}}
\end{figure}

Figure 2(i - l) show space-time diagrams of the curved trajectories of the downflow blobs. We fit the trajectories with a linear function, The slope represents the average plane-of-sky velocity. Acceleration is hard to see in the diagrams, indicating nearly constant plane-of-sky velocity of the blobs. The upper thread has larger velocity than the lower thread, which implies that outer threads holding faster downflows than inner threads. We further estimate the real velocity and acceleration in \S{3.2}.

Figure 3 plots the statistical distribution of all the detectable blobs velocities within the NST FOV. We detected 81 blobs in total from four hours before until ten minutes before the eruption. It is not easy to trace the blobs during the flare eruption due to dramatic intensity enhancement. The downflow velocity ranges from 28 to 63 km s$^{-1}$, with a mean velocity of 45 km s$^{-1}$. Outer threads host faster downflows than inner threads. Low velocity part is from 28 to 40 km s$^{-1}$, within the range of other filament plane-of-sky velocities (see \citet{2013SoPh..288..191J}; \citet{2014IAUS..300...69P}; \citet{2015ASSL..415...79K}). The materials with high velocity in the range of 50 to 63 km s$^{-1}$ occupied along the outer threads. They may suggest that the high velocities are augmented by the general upward motion associated with the slow-rise phase. The velocity of downflows does not vary dramatically during the four hours before the eruption from large-scale perspective. The small-scale inconstancy of velocity distribution can help explain the merge and split of those dark materials when they are moving along designated paths. 

\begin{figure}
\epsscale{0.6}
\figurenum{5}
\plotone{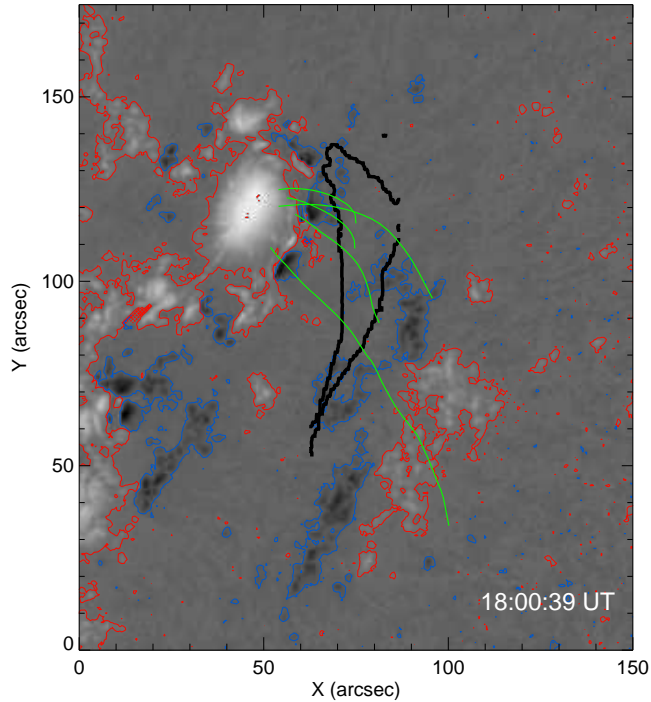}
\caption{SDO HMI LOS magnetogram with contours of $\pm 150$ G. The black contours indicate the locations of the filament obtained from an H$\alpha$ intensity image. The green lines show the extrapolated field lines above the penumbra. \label{fig:general}}
\end{figure}

\begin{figure}
\figurenum{6}
\gridline{ \fig{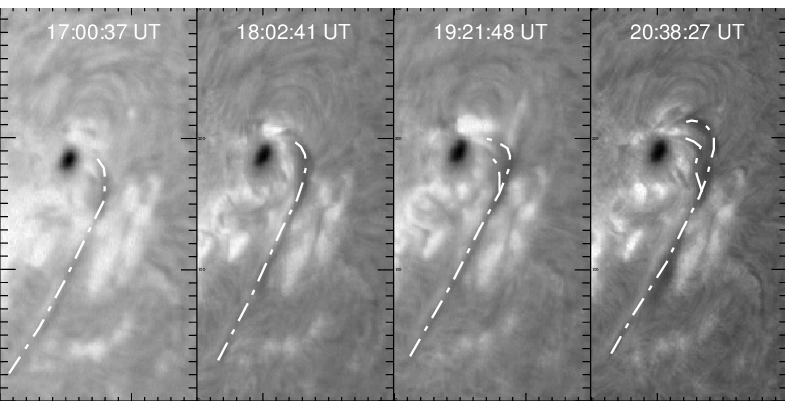}{0.8\textwidth}{}
}
\gridline{ \fig{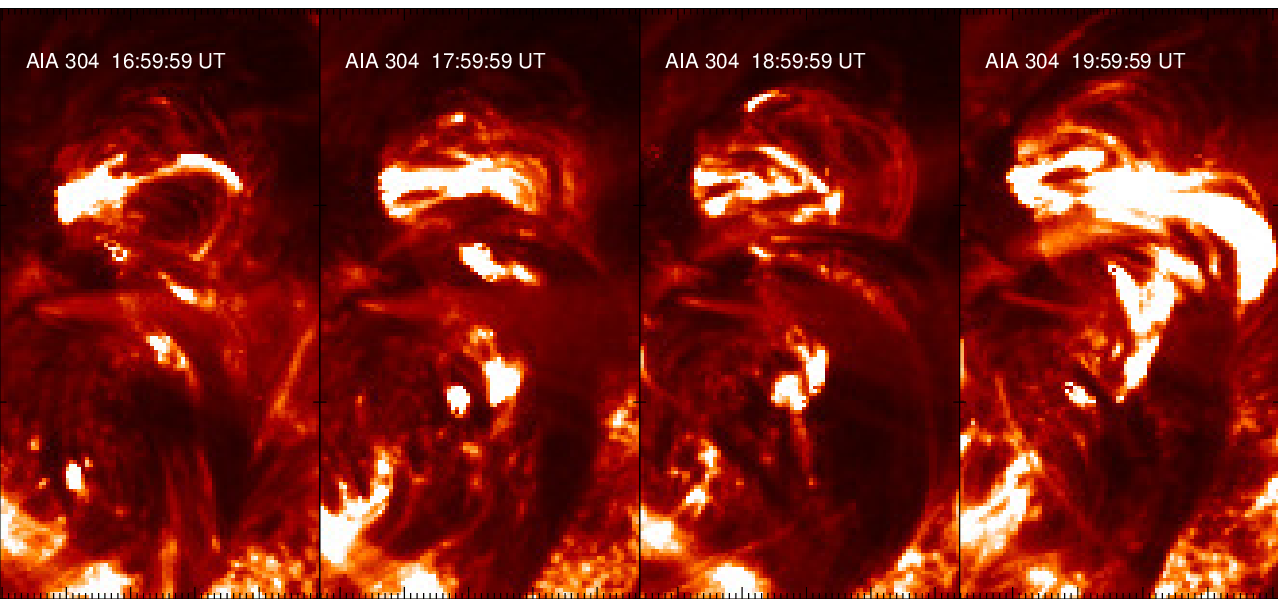}{0.8\textwidth}{}
}

\caption{Upper panels: H$\alpha$ intensity image taken from 17:00:37 UT - 20:38:27 UT before the flare eruption that occurred. They show the temporal evolution of the filament where spine is marked by the dashed lines. Bottom panels: AIA 304 $\AA$ images taken from 17:00 UT - 20:00 UT.  \label{fig:general}}
\end{figure}

\subsection{Filament foot real velocity} \label{subsec:tables}
We use NLFFF extrapolation to deduce local magnetic field. The spatial scale of NLFFF is 2.25'', much coarser than that of high resolution H$\alpha$ images. In Figure 4 extrapolated field lines at 18:48:00 UT are selected to fit ballistic trajectories of downflows. The ballistic trajectory matches extrapolated fields within the FOV of the high resolution H$\alpha$ data. We see many dense materials moving along the field lines. We chose these five NLFFF lines because their footpoints are close to Brightening Zones (BZs) and hence can help us identify whether these downflow threads are actually  landing upon BZs. The peak height of field lines ranges from 5 Mm to 15 Mm, which could be the lower limit of the altitude of the filament spine. Assuming the mass motion follows NLFFF lines, we find that the de-projected velocity is in a range of 40--80 km s$^{-1}$, with a mean of 56 km s$^{-1}$. Both inner threads and outer threads have faster downflows than that in quiescent phase \citep{2015ASSL..415...79K,2014IAUS..300...69P}. Due to high level of activity before or during eruption,Local field topology may lead to the nonuniform distribution of the velocity. Figure 5 shows the location of the filament over SDO/HMI LOS magnetogram. The spine of the filament lies along the PILs while the materials at the end moves along the magnetic field lines tending to cross the PILs. This suggests that the dense materials may accumulate at the magnetic field dips above the PIL, and then slide off along the filament end. 

To calculate total kinetic energy of the downflows directly flowing into BZs, we assume that the trajectory is a tilted axis-symmetric cylinder when flows are close to surface. Electron density of the filament is 10$^{9}$ - 10$^{11}$ cm$^{-3}$ \citep{1985SoPh..100..415H,2010SSRv..151..243L}, here we choose the lower limit of the electron density n$\textsubscript{e}$=8$\times$10$^{9}$cm$^{-3}$ for erupting filaments, and proton mass m$\textsubscript{p}$=1.67$\times$10$^{-24}$g. Mass density $\ell$\textsubscript{0}=n$\textsubscript{e}$m$\textsubscript{p}$= 1.33$\times$10\textsuperscript{-14} g cm\textsuperscript{-3}. A = 1.32 Mm$^2$ was chosen from the lower limit of the BZ's area, which is determined by 95$\%$ of peak intensity. The kinetic energy per second (KE) of downflow reaching upon the BZ is estimated to be 5.6$\times$10\textsuperscript{21}$\sim$4.5$\times$10\textsuperscript{22} ergs. 5.6$\times$10\textsuperscript{21} ergs is regarded as the lower limit of the kinetic energy per second through the BZs.
\[KE=\frac{Av\ell_{0}\cdot v^{2}}{2}\]

\begin{figure}
\figurenum{7}
\gridline{\fig{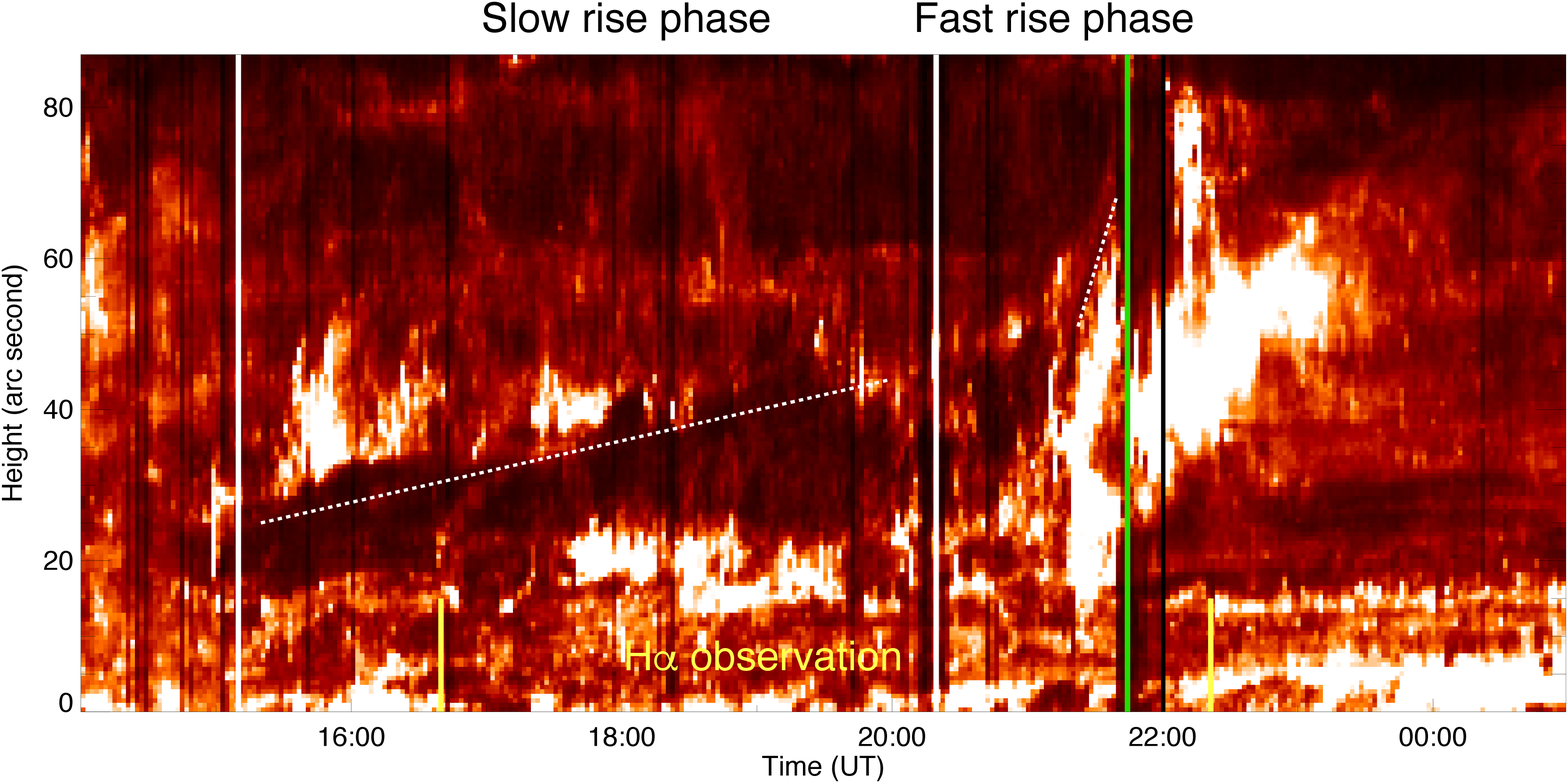}{0.7\textwidth}{}
              \fig{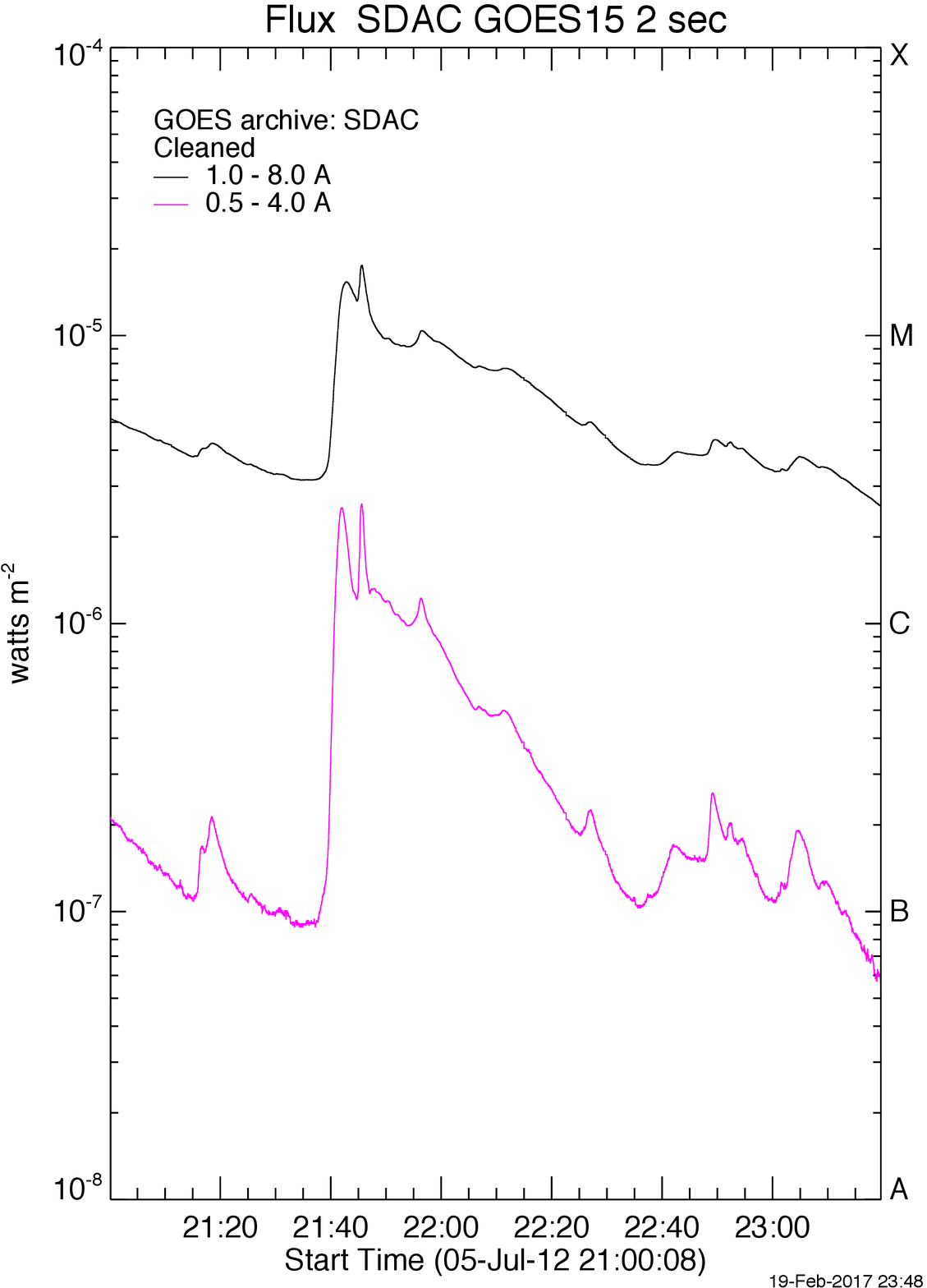}{0.3\textwidth}{}
}

\caption{Left panel: AIA 304 $\AA$ slit image. White lines show the slow and fast rise phases before eruption. White dash lines are fitted for the plane-of-sky velocities of the filament in the slow-rise phase and the end of the fast-rise phase respectively. Yellow lines show the duration of high-resolution H$\alpha$ observation, wherein the time interval from 16:41 UT to 20:51 UT is the effective duration for tracking the downflow blobs, the time interval from 16:41 UT to 22:22 UT is the overall observation. Green line represents the M1.6 flare peak time at 21:45 UT. Right panel: GOES X-ray flux, showing the M1.6 flare on 2012-07-05. \label{fig:general}}
\end{figure}

\subsection{Filament Drainage} \label{subsec:tables}
Figure 6 shows the temporal evolution of the filament before eruption. We can see one end of the filament anchored at the sunspot (AR 11515) ascending while the other end and the spine part keeping stationary. Arch-like expansion is one of the main procedures an entire pre-flare filament may undergo \citep{1963AJ.....68..290R}. In a case study by \citet{2015ApJ...806....9K}, a filament shows blue-shifted Doppler velocities of 1--10 km s$^{-1}$ at the pre-eruption phase. Since the downflow threads are confined to the magnetic fields, the expansion and ascent of the filament arcade lead to more radial magnetic field lines and downflow threads injection. Therefore, more dense materials could be poured downward back to lower chromospheric layer, resulting in drainage. 

The slit image shown in Figure 7 left panel is derived from SDO/AIA 304 $\AA$ images along the dashed line shown in Figure 1(d). The right panel shows the GOES X-ray flux of the M1.6 flare peak at 21:45 UT. The Filament started to ascend from $\sim$15:10 UT until the eruption at $\sim$21:30 UT. Slow-rise phase started from $\sim$15:10 UT with an average plane-of-sky velocity of 0.8 km s$^{-1}$ and the fast-rise phase started at $\sim$20:20 UT until the eruption with a final plane-of-sky velocity of 12 km s$^{-1}$. It is worth mentioning that the rising velocities are probably severely underestimated due to projection effects.  In a limb event where projection effects are less pronounced \citep{2015SoPh..290.1703M}, the average slow-rise and fast-rise velocities of a prominence are 2.1 and 106 km s$^{-1}$ respectively, both of which are higher than our results, 0.8 and 12 km s$^{-1}$. The high-resolution H$\alpha$ observation starts from 16:41:47 UT to 22:22:47 UT, and the effective duration for tracking downflow blobs starts from 16:41:47 UT to 20:41:47 UT, which mostly covers the slow-rise phase. Considering cross-sectional area A to be $\sim$300 Mm$^2$, we estimated the mass draining rate A$\cdot$$\ell$\textsubscript{0}$\cdot$v, which is 2.2$\cdot$10\textsuperscript{11}g$\cdot$s$^{-1}$ and 4.0$\cdot$10\textsuperscript{15}g for 5 hours continuous draining during the filament slow-rise phase. This is comparable to the filament clump mass $\sim$ 10$^{15}$g reported by \citet{2014ApJ...790..100B}. The draining rate is faster than $\sim$10$^{15}$g$\cdot$day$^{-1}$ in a non-eruptive filament studied by \citet{2012ApJ...745L..21L}. It drains away certain fraction of the filament mass that is required to confine the magnetic structure, thus may trigger the transition into fast-rise phase. As some models indicated \citep{2002ApJ...571..987F,2003ApJ...594.1060L,2005ApJS..159..288P}, filament drainage facilitates the eruption. The flare begins after the slow-to-fast rise phase transition. It  may rule out the flare as a potential trigger for the filament eruption. This is different from the event reported by \citet{2007ApJ...668..533N}, in which a C2.9 flare that occurred when the filament was close to the critical point for loss of equilibrium lead to the filament eruption. It should be noted that though the potential roles of the torus and kink instabilities are not discussed here, they can still be responsible for triggering the fast-rise phase. In such a way, the persistent downflows may make the filament rise further to meet the critical point for the instabilities.

%Drainage occurred since the beginning of the observation, so we are unable to identify which one contributes to the other one's onset. However, we believe the drainage comes with the slow-rise phase, and they both could contribute to the onset of fast-rise phase.

\subsection{Chromospheric brightening} \label{subsec:tables}

\begin{figure}
\figurenum{8}
\gridline{ \fig{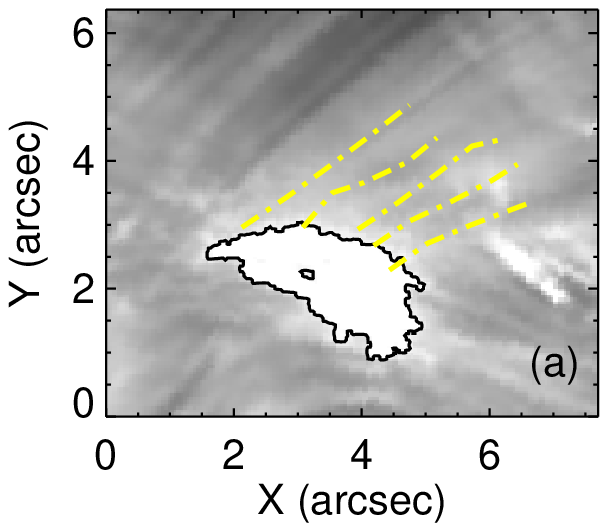}{0.33\textwidth}{}
          \fig{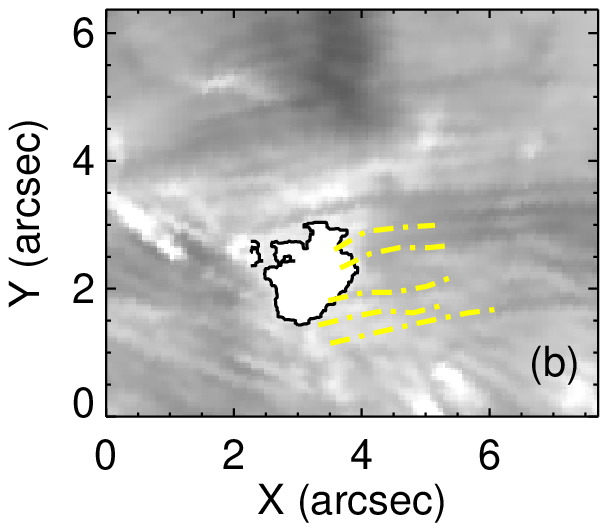}{0.33\textwidth}{}
           \fig{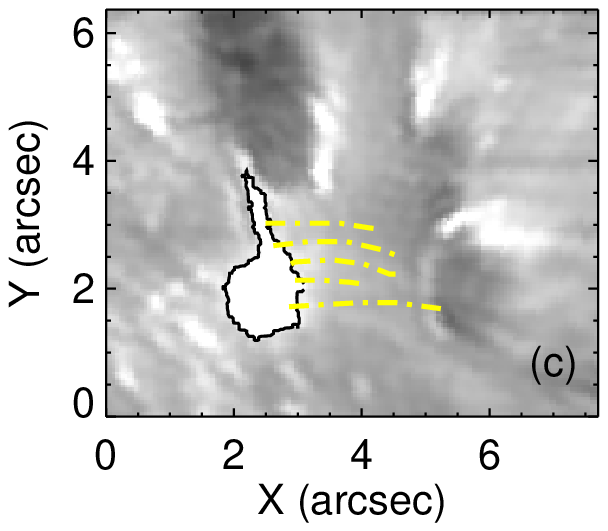}{0.33\textwidth}{}
          }
\caption{Three main brightening zones on chromosphere with the corresponding downflows marked in dashed lines. \label{fig:pyramid}}
\end{figure}

\begin{figure}
\figurenum{9}
\gridline{ \fig{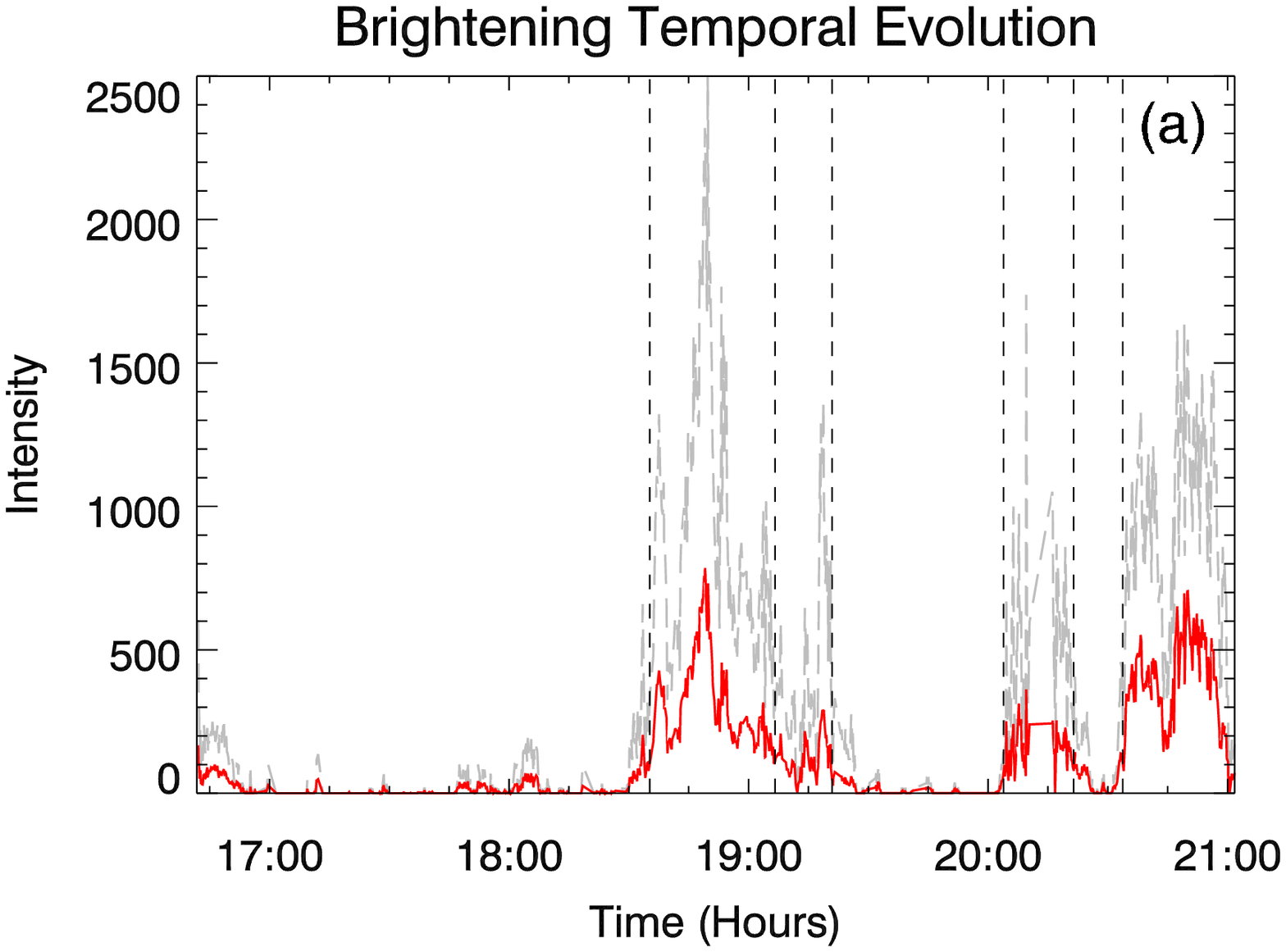}{0.33\textwidth}{}
          \fig{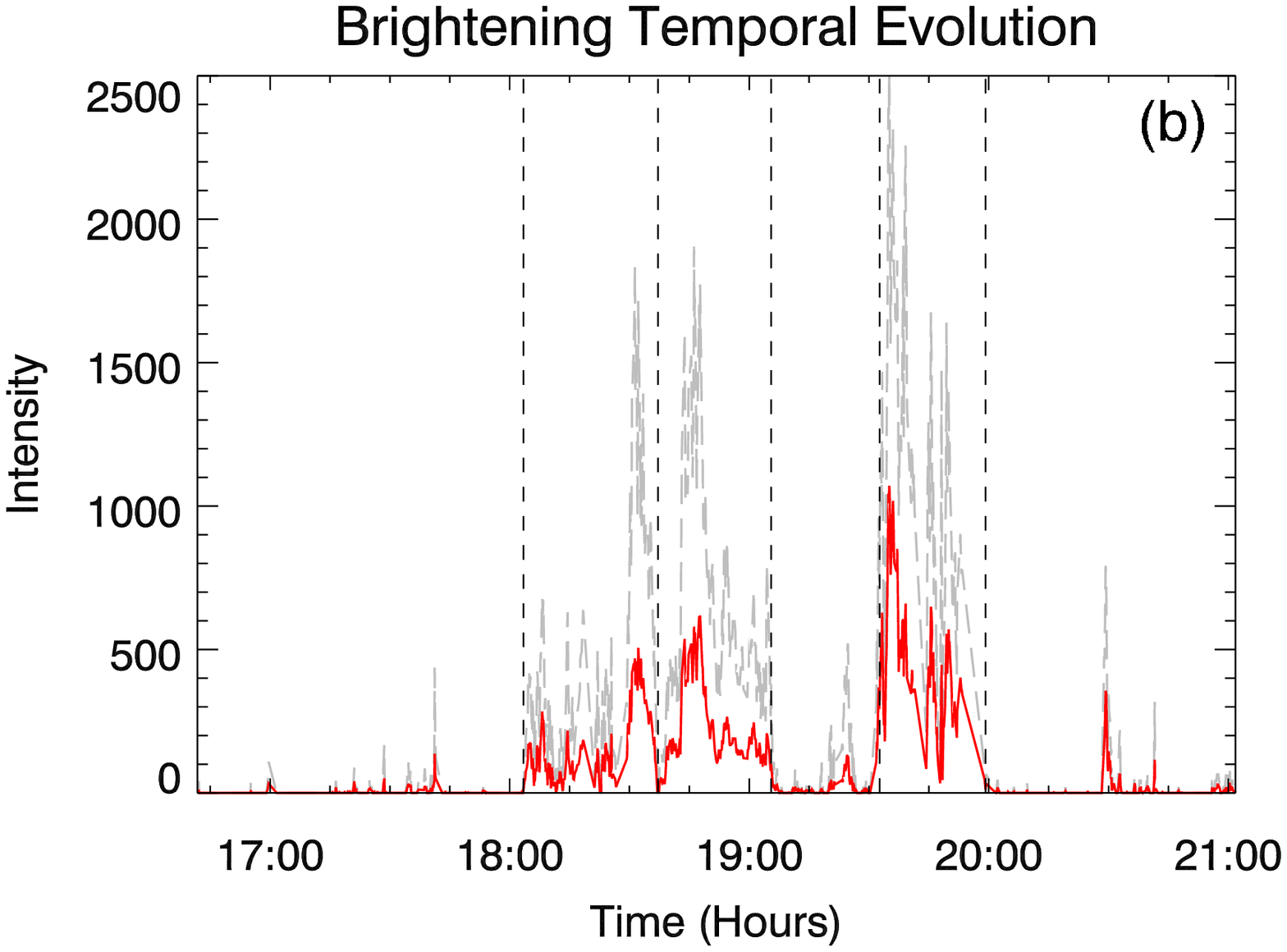}{0.33\textwidth}{}
          \fig{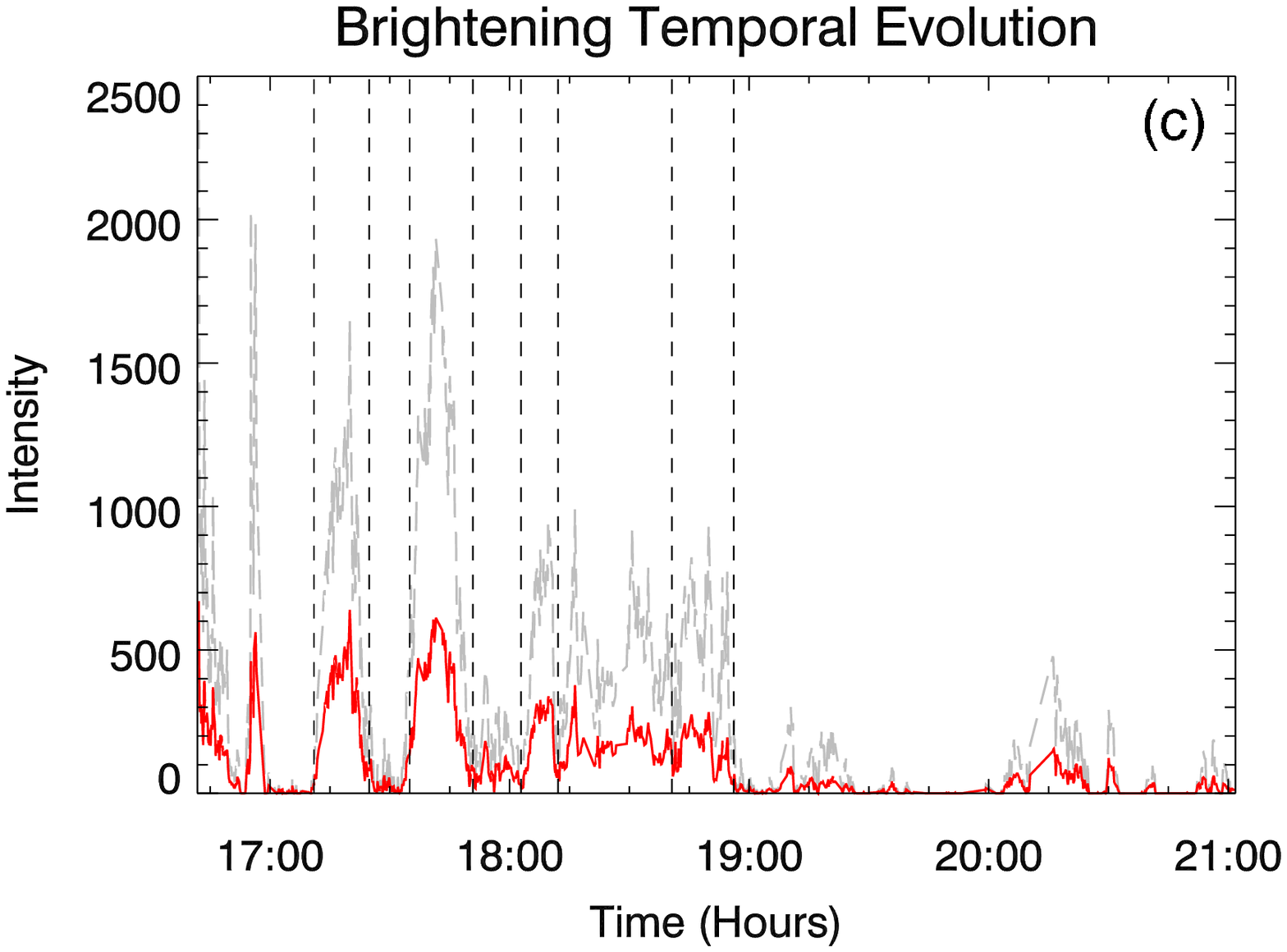}{0.33\textwidth}{}
          }
\caption{The temporal evolution of the average value of H$\alpha$ intensity averaged over BZs. Grey and red lines represent the total intensity before and after subtraction of background contrast respectively. Dashed lines mark each interval of BZs. \label{fig:pyramid}}
\end{figure}

\citet{2013ApJ...776L..12G} reported the chromospheric brightenings as the response to the impact of falling material from a partial filament eruption. It was proposed that the plasma compression is the dominant mechanism responsible for the observed brightenings. Here small-scale BZs are also observed associated with the filament downflows from 16:41:47 UT to 21:19:24 UT in the high-resolution observation. Three main BZs of characteristic length of $\sim$1.5$\arcsec$ are marked in Figure 1(a) and Figure 8. The BZs reside at the foot points of numerous downflow threads which are marked by dashed lines. The BZs may result from plasma heating due to the kinetic energy of the falling material. Unfortunately the BZs that are just barely resolved in AIA images are shaded by filament dark features, which makes it difficult to derive the radiated energy of the BZs from DEM. Figure 9 shows the temporal evolution of the total H$\alpha$ intensity over the area of BZs with/without subtracting background contrast. Background fluctuation does not affect the timing of brightenings. The size of these BZs kept relatively constant. In addition, these BZs recur successively, with an average lifetime of 20 min. 
%Radiation energy of the BZ is expected to be 10\textsuperscript{2} less than 10\textsuperscript{25}$\sim$10\textsuperscript{26} ergs based on emitting areas, which is given by \citet{2013ApJ...776L..12G} for a larger brightening event. Kinetic energy is comparable to radiation energy release when we also take minutes traveling time into consideration, i.e Kinetic energy is \textsuperscript{23}$\sim$10\textsuperscript{24} ergs for 3 minutes traveling. 
Both lifetime and size of these BZs are 1 order of magnitude greater than those of the brightening footpoints of post-flare downflows, as reported by \citet{2016NatSR...624319J}. 

\section{Conclusion} \label{sec:style}
To summarize, high-resolution NST H$\alpha$ observation of NOAA AR11515 revealed strong unidirectional downflows towards one end of the filament at the pre-eruption phase of the filament. 
%This type of mass flow is different from previously reported counter-streaming motion in a quiescent filament which maintains more "static". 
The ballistic trajectories of downflows keep steady except for a gradual ascent before the filament eruption. The extrapolated NLFFF lines near the AR end match well with the ballistic trajectories of downflow threads, which helps estimate the real velocities of the downflow threads. 
%Assuming that the mass motion follows NLFFF lines, we estimated the real velocities in a range of 40--80 km s$^{-1}$ with a mean of 56 km s$^{-1}$. The resulting kinetic energy per second corresponding to a BZ is 10\textsuperscript{21}$\sim$10\textsuperscript{22} erg. Comparable to radiation energy of a BZ \citep{2013ApJ...776L..12G}. 
Since our high-resolution observation started after the onset of the slow-rise phase, we are unable to identify whether filament drainage facilitate the onset of the slow-rise phase or vise versa. However, as the persistent drainage accompanied with the slow-rise phase as early as at least $\sim$4 hours before jumping into fast-rise phase, we believe that the drainage not only contributes to filament slow-to-fast-rise phase transition \citep*[e.g.][]{2014ApJ...790..100B,2014A&A...566A.148H}, but also maintains or even accelerates the slow-rise-phase. Slow-rise phase is often attributed to increasing twist in the flux rope \citep{2003ApJ...594.1060L,2002ApJ...571..987F}, which results in an upward force, that is essential for the eventual loss of equilibrium. Pre-eruptive drainage, accompanied with the slow-rise phase, could be regarded as a precursor to the eruption.
Successive brightenings were observed on the chromosphere when impacted by the falling material. The plasma heating due to the dissipation of kinetic energy into the chromosphere via compression is a more accountable mechanism for the observed brightenings. We speculate that downflows towards one filament footpoint result from the draining effect of the ascending filament prior to its eruption, and might be a precursor to eruptive events.

\acknowledgments

We thank Dr. Chang Liu for providing ideas to explain the observation, and Dr. Xin Chen for providing tracing method. Observations presented in this paper were mainly obtained with NST at Big Bear Solar Observatory, which is operated by New Jersey Institute of Technology. Data are courtesy of the BBSO and NASA/SDO teams. This work was partially supported by NASA grants NNX13AF76G, NNX16AF72G, NNX13AG13G, and NNX14AC12G, and by NSF grants AGS 1348513, 1408703 and 1620875. BBSO operation is supported by NJIT, US NSF AGS-1250818 and NASA NNX13AG14G, and NST operation is partly supported by the Korea Astronomy and Space Science Institute and Seoul National University, and by strategic priority research program of CAS with Grant No. XDB09000000.

\listofchanges

\end{document}